\documentclass[useAMS,usenatbib]{mn2e}

\usepackage{dsfont}
\usepackage{amsmath}
\usepackage{amssymb}
\usepackage{graphicx}
\usepackage{verbatim}
\usepackage{natbib}
\usepackage{eucal}
\usepackage{calligra}

\usepackage[usenames,dvipsnames]{color}

\usepackage{amsfonts}
\usepackage{color}
\usepackage[normalem]{ulem}
\usepackage[T1]{fontenc}

\DeclareMathAlphabet{\mathscr}{OT1}{pzc}%
                                 {m}{it}
                                 
\newcommand{\mnras}{MNRAS}

\newcommand{\apj}{ApJ}
\newcommand{\apjl}{ApJL}

\newcommand{\be}{\begin{equation}}
\newcommand{\ee}{\end{equation}}
\newcommand{\bes}{\begin{equation*}}
\newcommand{\ees}{\end{equation*}}
\newcommand{\bea}{\begin{eqnarray}}
\newcommand{\eea}{\end{eqnarray}}
\newcommand{\beas}{\begin{eqnarray*}}
\newcommand{\eeas}{\end{eqnarray*}}

\newcommand{\mpch}{\;{\rm Mpc}/h}

\newcommand{\zl}{z_{\rm L}}
\newcommand{\zs}{z_{\rm s}}

\newcommand{\sv}{s_{\rm v}}
\newcommand{\rv}{R_{\rm v}}
\newcommand{\rvm}{R^{(m)}_{\rm v}}
\newcommand{\thetav}{\theta_{\rm v}}
\newcommand{\rlos}{r_{\rm los}}

\newcommand{\fvol}{f_{\rm vol}}

\def\ave#1{\left\langle #1 \right\rangle}

\begin{document}

\title[Void Lensing]
{Lensing Measurements of the Mass Distribution in SDSS Voids }
\author[J. Clampitt \& B. Jain]
  {Joseph~Clampitt$^{1,}$\thanks{E-Mail: clampitt@sas.upenn.edu}, Bhuvnesh~Jain$^{1}$\\
  $^1$Department of Physics and Astronomy, Center for Particle Cosmology, \\
  University of Pennsylvania,
  209 S. 33rd St., Philadelphia, PA 19104, USA}

\maketitle

\begin{abstract}
We measure weak lensing mass profiles of voids from a volume-limited sample of SDSS Luminous Red Galaxies (LRGs). We find voids using an algorithm designed to maximize the lensing signal by dividing the survey volume into 2D slices, and then finding holes in this 2D distribution of LRGs. We perform a stacked shear measurement on about 20,000 voids with radii between $15-55 \mpch$ and redshifts between $0.16-0.37$. We measure the characteristic radial shear signal of voids with a signal-to-noise of 7. The mass profile corresponds to a fractional underdensity of about -0.4 inside the void radius and a slow  approach to the mean density indicating a partially compensated void structure. We compare our  measured shape and amplitude  with the predictions of Krause et al 2013. Voids in the galaxy distribution have been extensively modeled using simulations and measured in the SDSS. We discuss how the addition of void mass profiles can enable studies of galaxy formation and cosmology. 
\end{abstract}

\begin{keywords}
voids, weak lensing
\end{keywords}

%%%%%%%%%%%%%%%%%%%%%%%%%%%%
\section{Introduction}
%%%%%%%%%%%%%%%%%%%%%%%%%%%%

The first measurement of lensing from stacked galaxies was observed almost twenty years ago by \citet{bbs1996}. Since then, applications of this technique to the Sloan Digital Sky Survey (SDSS) have made stacked galaxy lensing an indispensable measure of galaxy halo masses, e.g., \citet{mhs2005} and \citet{sjs2009}. More recently, in \citet{cjt2014}, we measured the stacked lensing signal of filaments connecting neighboring Luminous Red Galaxies (LRGs).  We were able to study filament properties such as thickness by comparing to simulation results. Now, with the goal of obtaining such a measurement of voids, we construct a void catalog from holes in the LRG distribution of SDSS, measure the void tangential shear profile, and constrain their  density profiles.

There are many void finders in the literature, all differing in implementation and the resulting set of voids found. \citet{cpf2008} makes a comparison of 13 algorithms. In recent years, methods involving a Voronoi tessellation coupled with a watershed transform have become popular \citep{pvj2007, n2008, lw2012}. These methods have also been successfully applied to data, yielding void catalogs from surveys such as SDSS \citep{slw2012}. A lensing analysis of the \citet{slw2012} catalog was carried out by \citet{mss2013}. However, despite careful attention to details of the shear measurement, the small number of voids in the catalog was likely a factor in the marginal detection significance.

Recent work has studied in more detail the properties of dark matter voids in simulations.  \citet{hsw2014} found that previous fits to simulation density profiles were too simple and provide fitting formulae with parameters that can be adapted to voids with a range of sizes. \citet{slw2013} and \citet{slw2014} have worked to connect the theory of voids found in the dark matter to those found in galaxies by using Halo Occupation Distribution models to mimic realistic surveys. Excursion set work has focused on providing semi-analytical models of void abundances \citep{sv2004, pls2012}, as well as connecting these models to void counts from simulations \citep{jlh2013}.

Once void catalogs are constructed, they have numerous other applications. \citet{hvp2012} used a different void finder \citep{hv2002, pvh2012} to study the photometric properties of void galaxies. They find that void galaxies are bluer than those in higher density environments, but do not vary much within the void itself. Cosmological probes such as the Alcock-Paczynski test \citep{lw2012, slw2012b} and void-galaxy correlations \citep{hws2014} have been proposed. Finally, voids also provide a sensitive test of some modified gravity theories \citep{lzk2012, ccl2013, cpl14a, cpl14b, lcc15, bcl15}.

Section 2 describes our basic void-finding algorithm, as well as our cuts to select a subsample useful for lensing. Section 3 explains our weak lensing measurement, null tests, and expected signal-to-noise. Section 4 presents our results on void density profiles, including both a fitted model and model-independent statements. Finally, Section 5 summarizes our results, caveats, and directions for future work.

%%%%%%%%%%%%%%%%%%%%%%%%%%%%
\section{Void finding}
%%%%%%%%%%%%%%%%%%%%%%%%%%%%

In the following, we describe how void centers and radii are determined for voids in the galaxy distribution.
We then explain quality cuts designed to remove galaxy voids that are unlikely to correspond to dark matter underdensities.

%%%%%%%%%%%%%%%%
\subsection{Locating Galaxy Voids}
\label{sec:algorithm}
%%%%%%%%%%%%%%%%

As void tracers, we use the volume limited subsample of the SDSS DR7-Full LRG catalog of \citet{kbs2010}, which contains $\sim 66,500$ LRGs between $0.16 < z < 0.37$.
We begin by cutting the volume probed by LRGs into slices of comoving thickness between 20 and 100 $\mpch$, in steps of $10 \mpch$.
We then treat each slice with its set of LRGs as a 2D spherical surface, and look for galaxy voids on that surface.
Before describing our void finder in detail, we point out the close analogy between our algorithm and a simpler method.
One way to find voids is by passing 2D tophat filters of various radii $\thetav$ over the density field and recording the coordinates and filter sizes which contain no galaxies.
The resulting circles on the sky would greatly oversample the area, but smaller circles which fell completely within larger ones could be removed.
The remainder would constitute a catalog of potential void centers with radii $\thetav$.
The output of such an algorithm would be qualitatively similar to ours.
However, our method does a better job of assigning area elements uniquely to each void, and it is computationally more efficient since coordinates that are near LRGs never have to be considered at all.

%%%%%%%%%%%%%%%%%%%
\begin{figure}
\centering
\resizebox{85mm}{!}{\includegraphics{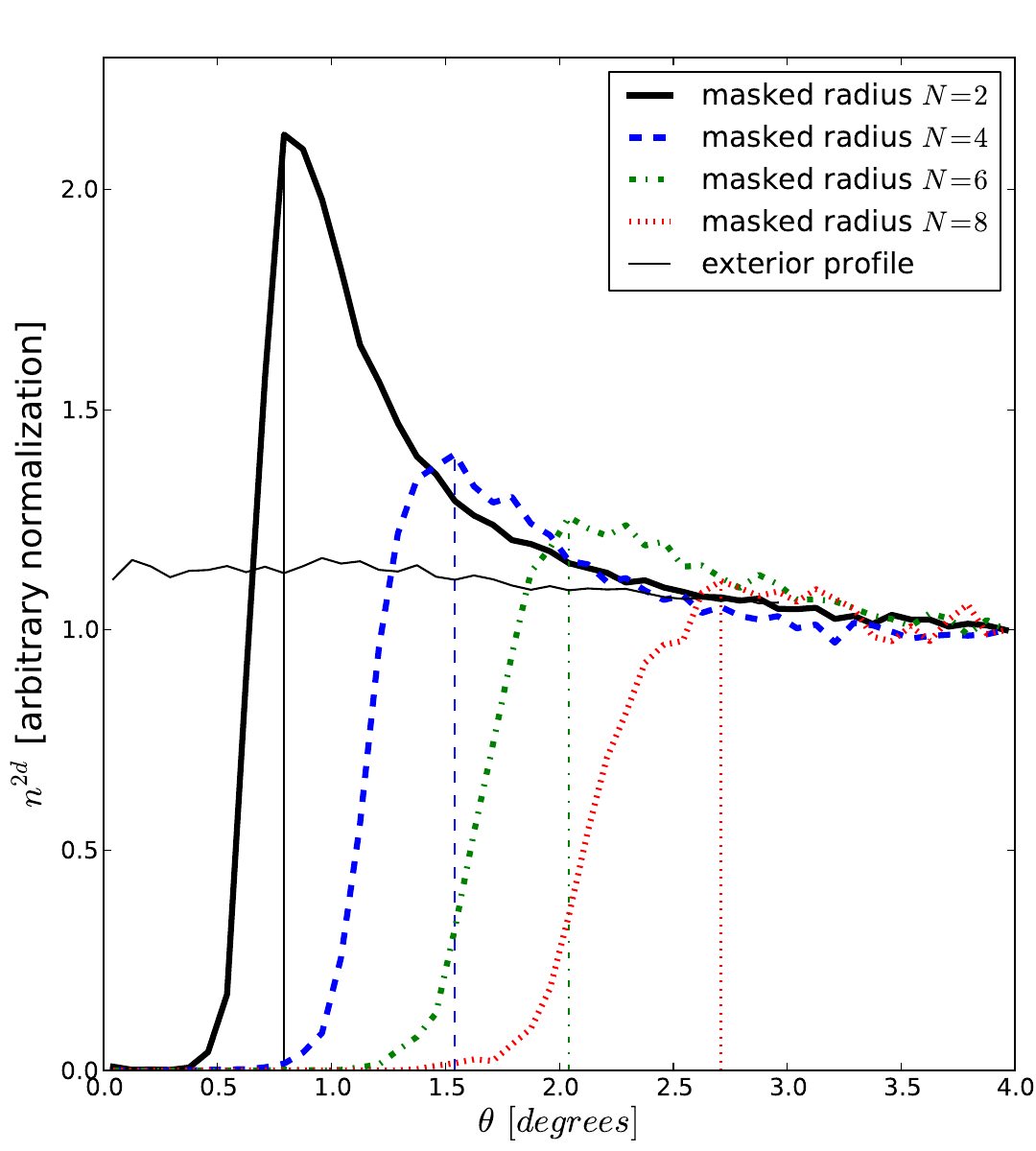}}
\caption{The 2D LRG density within the void slices, stacked over all galaxy voids found for a particular masking radius $N$.
The voids with smallest transverse size show a prominent ridge: we take the angular scale of the ridge maximum as the void radius $\thetav$.
The average surface density of LRGs outside the void slice is shown for all voids used in our lensing measurement (thin solid line).
The density of LRGs outside the slice is roughly the same for all radii, showing that on average our voids fit snugly within the 2D slices.
}
\label{fig:lrg-dens}
\end{figure}
%%%%%%%%%%%%%%%%%%%

Around each LRG location we mask out a region of radius $N$ pixels, consisting of all HEALpix\footnote{http://healpix.sf.net} \citep{ghb05} pixels that are at most $N$ pixels away from the LRG, moving in any direction.
We then group the unmasked area into contiguous regions, and any unmasked region that contains fewer than $N_{\rm threshold}$ pixels is counted as a galaxy void and completely masked out.
Applying this threshold ensures that as a set of unmasked pixels shrinks as the masked radius $N$ grows, the unmasked pixels are not counted as a void for multiple $N$.
Experimenting with other galaxy tracers besides LRGs, we found that the parameter $N_{\rm side}$ needs to be larger (resulting in a finer pixel grid) for denser galaxy tracers before the void finder converges.
For this LRG sample, a HEALpix resolution of $N_{\rm side}=256$ ensured convergence.
The mapping between masked radius $N$ and angle on the sky clearly depends on the resolution $N_{\rm side}$, as does $N_{\rm threshold}$.
For our galaxy sample and resolution, the results are not sensitive to the precise value of $N_{\rm threshold}$ within the range $\sim 10-30$.
For the masked pixel radii $N$ we started at 1 and moved to higher integers: by $N \sim 10$ most of the survey pixels were masked out and virtually no new voids were found.

%%%%%%%%%%%%%%%%
\subsection{Assigning Void Radii}
%%%%%%%%%%%%%%%%

The next step involves assigning to each galaxy void a radius on the sky and in the line-of-sight direction.
We begin by mapping each masked pixel radius $N$ to a specific angular void radius $\thetav$.
This is done empirically by calculating the average 2D LRG density around the galaxy void centers.
The result is shown in Fig.~\ref{fig:lrg-dens}.
Note the clear ridge in the galaxy profile that becomes less prominent for larger voids.
This trend is qualitatively consistent with the results of other void finders, e.g., the void profiles of \citet{hsw2014}.
We take the maximum of this ridge of galaxies as the void radius $\thetav$.
This angle is then converted to a comoving radius in $\mpch$ according to 
$\rv = \rlos \, \thetav$,
where $\rlos$ is the comoving distance to the void.
Later, in Sec.~\ref{sec:model}, we compare $\rv$ (which is determined from visible galaxies) to the dark matter void radius fit from the weak lensing data.

\begin{figure}
\centering
\resizebox{85mm}{!}{\includegraphics{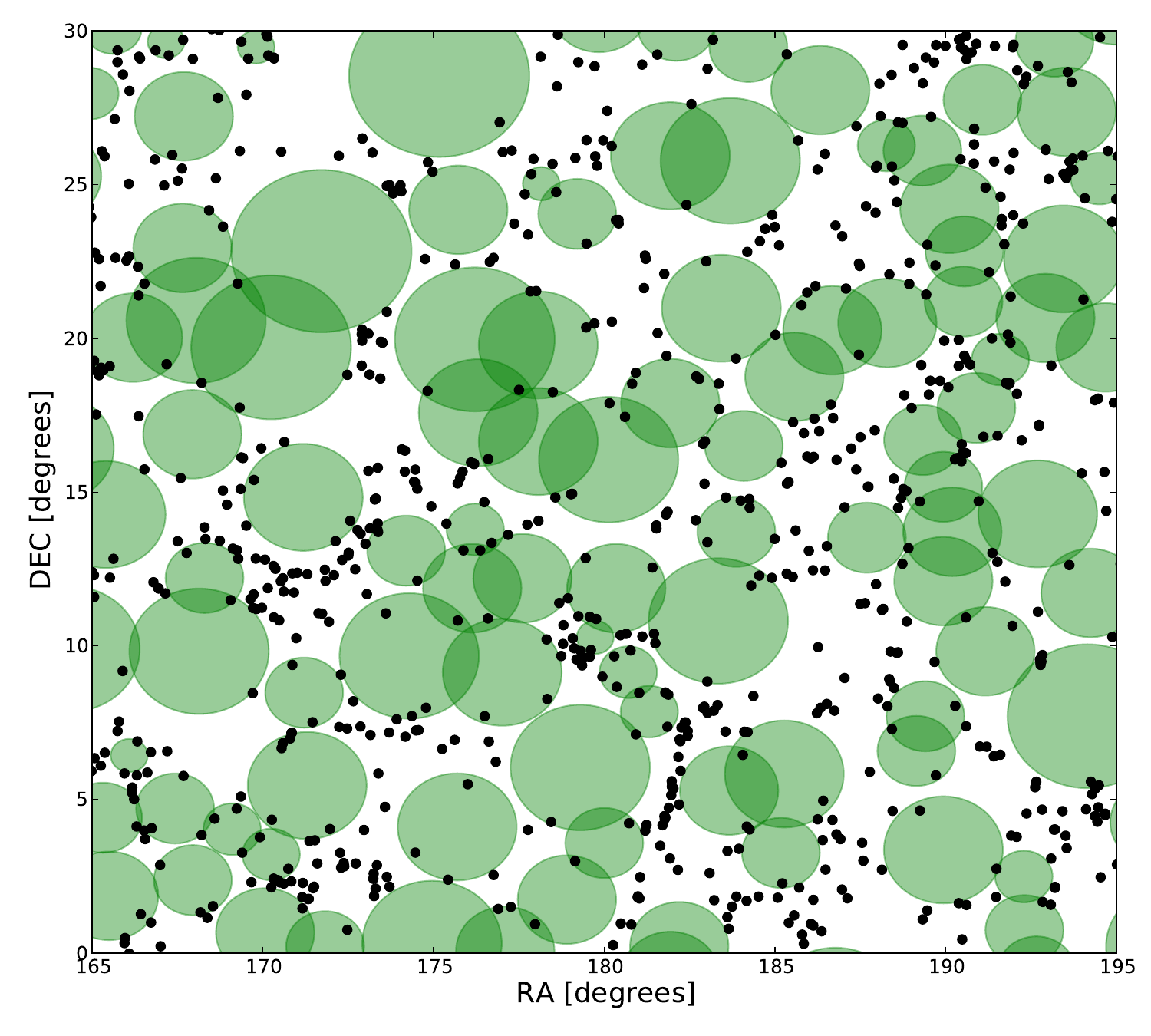}}
\caption{Slice thickness of $60 \mpch$, corresponding to voids with line of sight radius $\sv = 30 \mpch$. The black points show LRGs found within this slice. This is an intermediate redshift ($z = 0.24$) slice with typical volume and 2D LRG density. The green shaded circles show the output of our void finder. Many of the large gaps that are not identified as voids in this slice will be identified when sampling using a thicker or thinner slice.}
\label{fig:mid-z}
\end{figure}

In Fig.~\ref{fig:mid-z} we show an example of the galaxy voids found in a slice centered at $z = 0.24$ and with thickness $60 \mpch$.
The black points show LRGs that fall within this particular 2D slice, while the shaded green circles show the voids identified by our finder.
The purity of our sample is quite high, in the sense that the voids identified are all very empty of LRGs, with perhaps a few being found on the edges of voids.
In addition, the algorithm is able to find voids of a wide range of sizes.
For example, both the large void at (RA, DEC) = (172, 23) degrees with radius $\thetav \sim 2.5$ degrees,
and the very small void at (RA, DEC) = (178, 25) degrees with radius $\thetav \sim 0.5$ degree, are identified.
On the other hand, the algorithm does somewhat worse on completeness within the slice.
Consider the point (RA, DEC) = (188, 10) degrees, where by eye we identify a circular void with radius $\thetav \sim 1$ degree while the algorithm does not find any void.
Note however, that our void finder involves sampling the same survey volume multiple times with varying slice sizes.
Empty spaces that are missed in one slice will often be filled in when the slice thickness is very varied.
See Sec.~\ref{sec:volume} for more on how voids found in various slices are combined into a single catalog.

%%%%%%%%%%%%%%%%%%%
%%%%%%%%%%%%%%%%%%%

Having assigned the void radii in the plane of the sky, we now attempt to determine their sizes in the line-of-sight direction.
If our LRG voids fit snugly within the 2D slice in which each was identified, a natural choice is to use the slice thickness for the voids' line-of-sight size.
However, since the slice center and thickness are arbitrary and we have only used LRGs within the slice to determine the void location, it is not clear whether the galaxy void extends even farther.
To test this, we compare the LRG density within void slices to the density just outside; the result is shown in Fig.~\ref{fig:lrg-dens}.
At large radii $\sim 4$ degrees the LRG density both inside and outside the slice has leveled off to the mean.
Moving inwards towards the void center the two curves diverge: the galaxy ridge and then decrement are obvious within the slice, while outside the slice the galaxy density stays level near the cosmic mean.
Thus, on average our voids do fit snugly within each 2D slice, and we use the slice thickness for our void line-of-sight size.
For convenience and easy comparison to the plane-of-the-sky radius $\rv$, we define $\sv$ to be half the slice thickness.

%%%%%%%%%%%%%%%%
\subsection{Quality Cuts}
\label{sec:volume}
%%%%%%%%%%%%%%%%

\begin{figure*}
\centering
\resizebox{180mm}{!}{\includegraphics{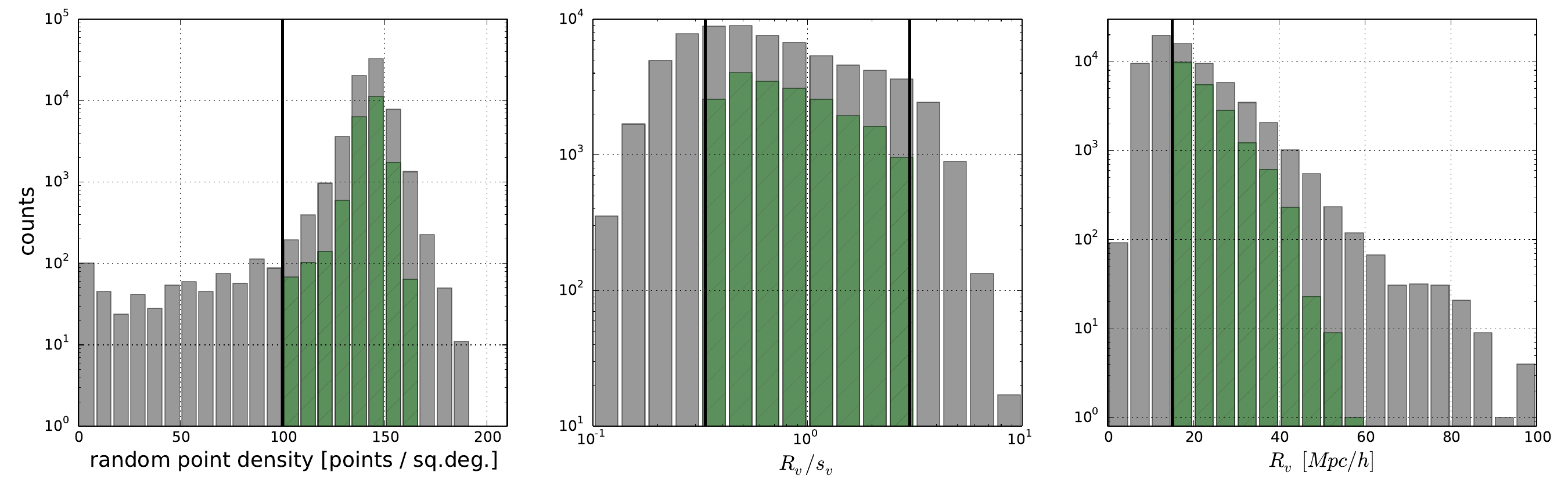}}
\caption{
({\it Left panel}): Distribution of density of random points within $\rv$ for our voids, both before (solid gray histogram) and after (hatched green histogram) quality cuts are applied. Fake voids which appear near survey masks and edges are eliminated with a minimum cut on this quantity.
({\it Center panel}): The ratio of the voids' projected size to their line-of-sight size $\rv/\sv$. Since voids should have axis ratios that do not differ to much from unity, we require our voids to have $1/3 < \rv/\sv < 3$.
({\it Right panel}): Distribution of projected void radii, which falls quickly with increasing size. We remove the smallest voids which are smaller than the mean interparticle spacing of the sparse LRG sample, requiring $\rv < 15 \mpch$.
These cuts, plus those on volume overlap (see text for details) trim the voids down to a catalog of $\sim 19,000$ objects.
}
\label{fig:analyze}
\end{figure*}

Having found a large set of galaxy voids numbering $\sim 68,000$ objects, we next remove those which are not likely to be dark matter voids. These include fake voids due to the survey mask and edges, chance alignments of LRGs in the projection, voids much smaller than the LRG mean interparticle spacing, and multiple detections of the same voids in overlapping slices.
First, an unusually high number of galaxy voids will be found at the survey edges and in regions where the LRG coverage is incomplete due to masking. In order to remove such spurious voids, we use the LRG random catalog from \citet{kbs2010}, which has $\sim 16$ times as many objects as real LRGs. For each void, we find the density of random points inside its angular radius $\thetav$. The histogram of densities is shown in the left panel of Fig.~\ref{fig:analyze}. The distribution is tightly peaked at 150 points/deg$^2$, with the densest voids having up to 200 points/deg$^2$. On the low-density end, there is a long tail stretching all the way to zero due to fake voids formed from unobserved regions. We remove the few hundred objects with density less than 100 points/deg$^2$ in this tail.

In addition, objects with very small axis ratios are likely to be chance alignments of holes in the sparse LRG sample.
Like halos, voids are expected to have axis ratios which are $\sim 1$ \citep{sv2004, pvj2008, ps2013}.
For example, \citet{pvj2008} find axis ratios of order c:b:a $\sim$ 0.5:0.7:1.0 using voids found in simulations.
Very long thin objects or flat ``pancake'' shapes are unlikely to have formed from a single negative fluctuation in the gaussian density field at early times.
We remove these by placing a cut on the line-of-sight and transverse axis ratio, requiring
$\sv / 3 < \rv < 3 \sv$.
The center panel of Fig.~\ref{fig:analyze} shows the distribution of the void size ratio, $\rv / \sv$, and the vertical lines display this cut.
Another type of chance hole involves those that are much smaller than the average interparticle spacing of LRGs, which for this sample is $\sim 20 \mpch$ \citep{kbs2010}.
Given that (i) LRGs are highly clustered, and (ii) most of our voids are larger in the line-of-sight direction (Fig.~\ref{fig:analyze}, middle panel), we put a minimum cut of $15 \mpch$ on the transverse radius $\rv$.
The distribution of void radii $\rv$ are shown in the right panel of Fig.~\ref{fig:analyze}.
Note also that the distribution of projected void radii (Fig.~\ref{fig:analyze}) is seen to fall rapidly with radius: this trend is qualitatively consistent with the results of previously mentioned void finders.

%%%%%%%%%%%%%%%%%%%
%%%%%%%%%%%%%%%%%%%

Finally, due to oversampling the same volume with many 2D slices of varying thickness, we sometimes find duplicate voids.
These duplicates have nearly identical center coordinates, but differing sizes along the line-of-sight.
We remove them by assigning a volume
$V = 2\sv \times \pi \rv^2$
to each void and then calculating the fractional volume overlap $\fvol$ with the void's nearest larger neighbor.
Such duplicates will have $\fvol = 1$, so we require $\fvol < 0.9$ for our fiducial void sample.
After applying all the above quality cuts, the resulting distributions of void properties are shown in Fig.~\ref{fig:analyze}.
The distribution of void sizes falls rapidly with increasing radius, with the largest remaining void measuring $\sim 56 \mpch$ in projected radius.

In addition to removing duplicate voids, the quanitity $\fvol$ is useful for studying the effects of allowing sub-voids and overlapping voids in our void sample.
The question of whether or not to allow voids to overlap varies substantially within the literature.
For example, ZOBOV \citep{n2008} outputs the entire void hierarchy, including all sub-voids that are nestled inside larger, more significant voids.
Other void finders do not return any such subvoids and allow no overlap between nearby voids, e.g., as done in \citet{jlh2013} for the purpose of comparison to excursion set void number functions.
Thus, it is interesting to study a second sample with $\fvol < 0.5$, for which the total number of voids is reduced by about a factor of 2 to $\sim 10,000$.
Interestingly however, our measured density profiles are not sensitive to the change in the $\fvol$ cutoff: the shift in the parameter contours for this sample is less than $1\sigma$, as will be shown in Fig.~\ref{fig:contour}.
The volume fraction in our fiducial void sample is about 0.7, whereas requiring $\fvol < 0.5$ the volume fraction in voids drops to $\sim 0.45$.
These values are comparable to those from other void finders in SDSS data: \citet{pvh2012} found volume fractions of about $0.6$, while \citet{slw2012} found values ranging from 0.12 to 0.43 depending on the void tracer.

%%%%%%%%%%%%%%%%%%%%%%%%%%%%
\section{Lensing Measurement}
%%%%%%%%%%%%%%%%%%%%%%%%%%%%

%%%%%%%%%%%%%%%%%%%
\begin{figure*}
\centering
\resizebox{80mm}{!}{\includegraphics{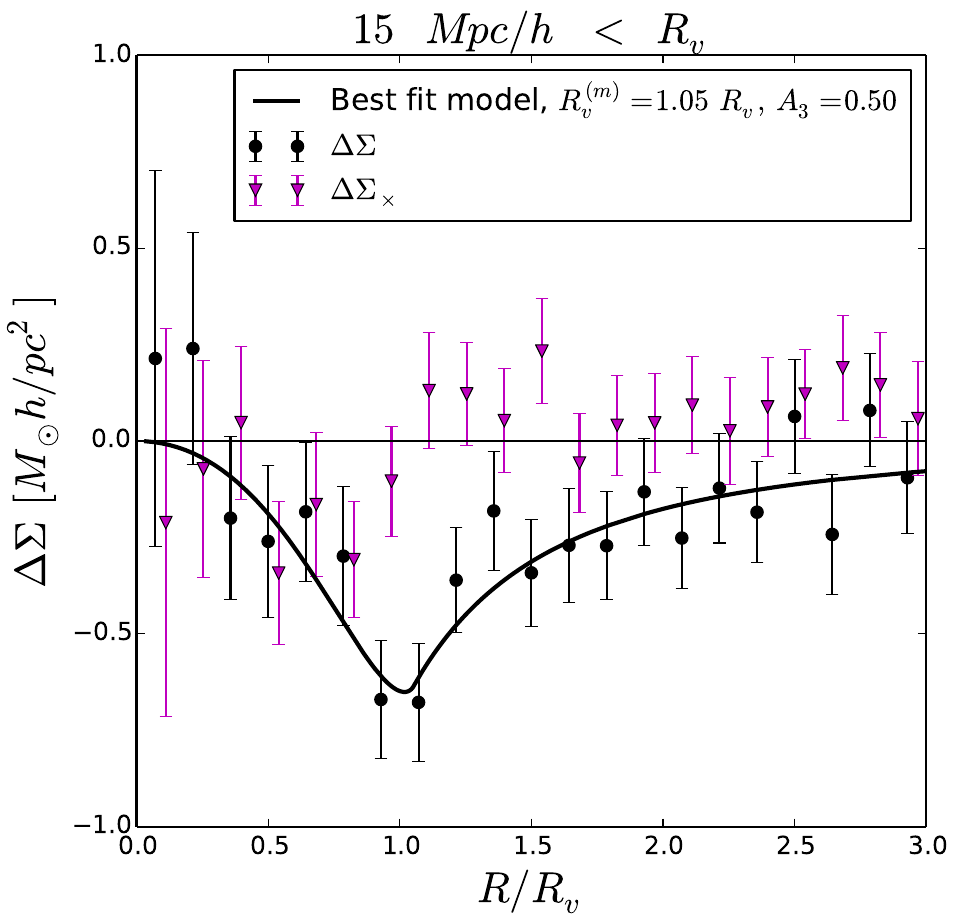}}
\resizebox{82mm}{!}{\includegraphics{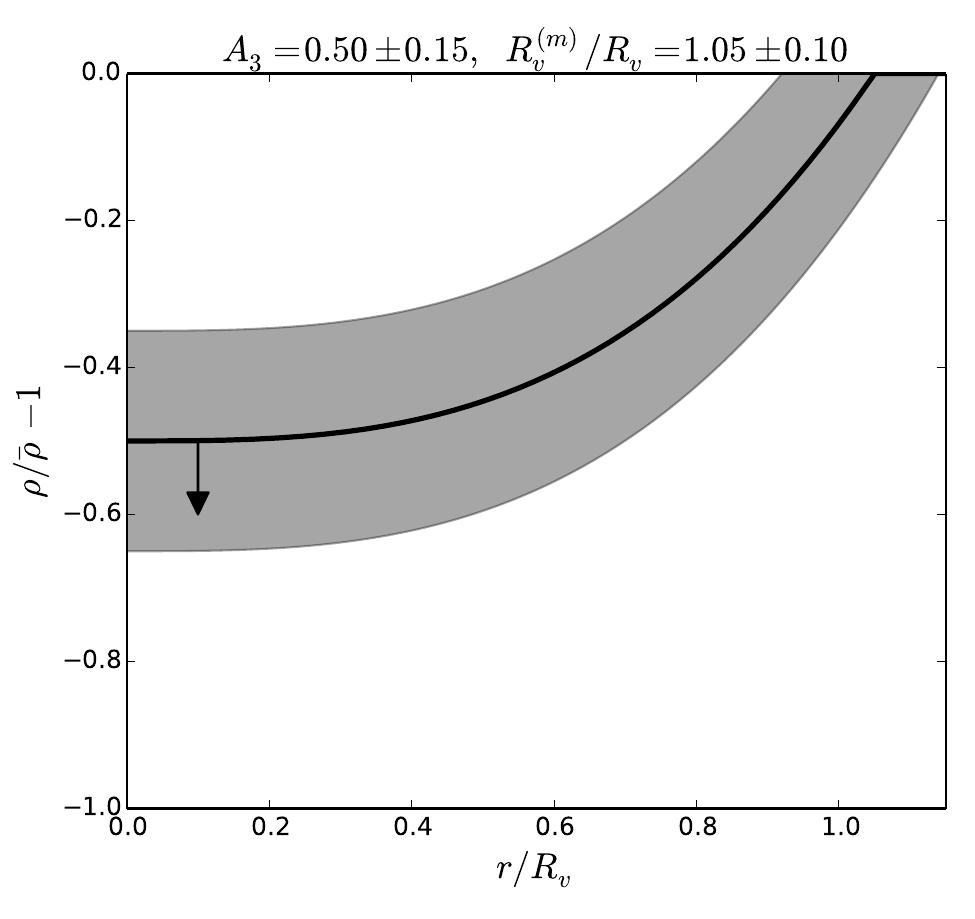}} \\
\caption{The left panel shows our measurement of the tangential shear (black circles) and cross-component (magenta triangles) around our void centers, stacked in units of $\rv$. Our best-fit model (solid line) has $\rvm = 1.05 \, \rv$ and $A_3 = 0.50$. Our estimated 3d density profile is shown in the right panel, along with the estimated 1$\sigma$ uncertainty. The arrow gives a sense of our model independent estimates, which prefer a lower central density (by up to $0.1\bar{\rho}$) than allowed by our model.
}
\label{fig:gammat}
\end{figure*}
%%%%%%%%%%%%%%%%%%%

%%%%%%%%%%%%%%%%%%%
\begin{figure*}
\centering
\resizebox{180mm}{!}{\includegraphics{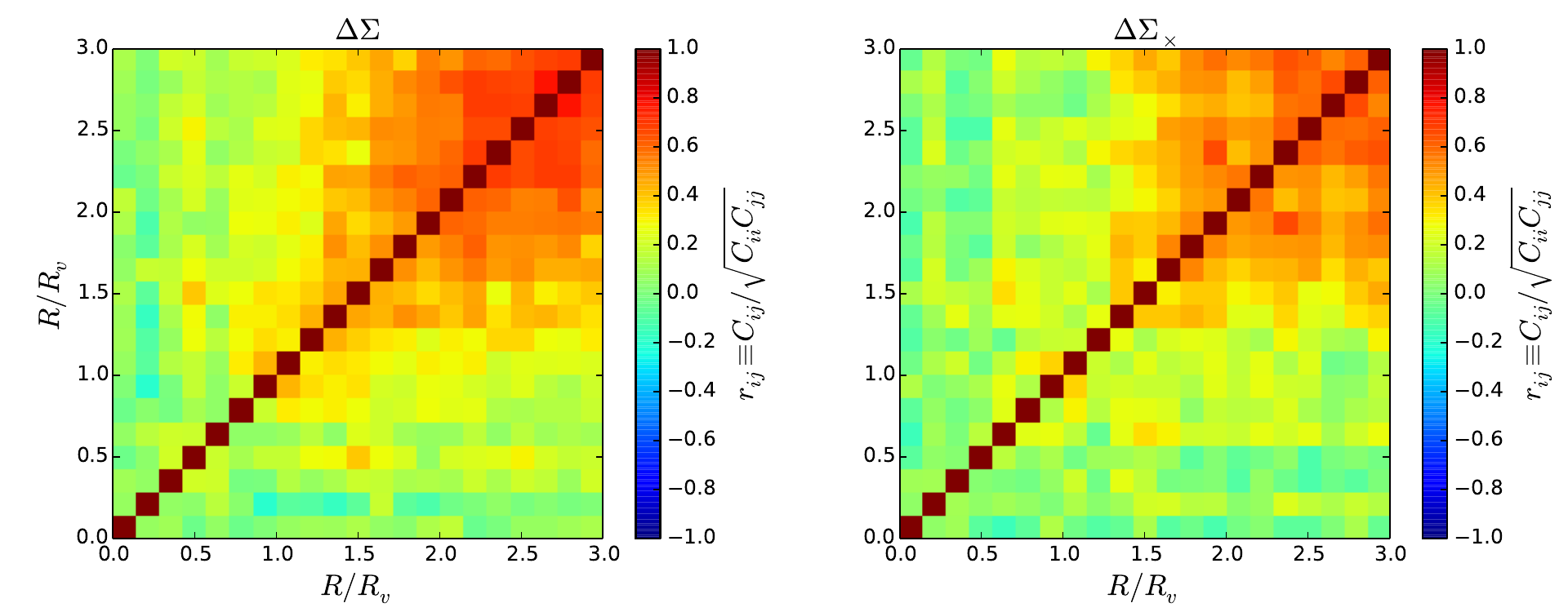}}
\caption{Our covariance matrices for $\Delta\Sigma$ ({\it left panel}) and $\Delta\Sigma_\times$ ({\it right panel}). Off diagonal correlations are significant beyond $2\rv$, since source galaxies in these bins are shared among multiple voids.}
\label{fig:cov-gammat}
\end{figure*}
%%%%%%%%%%%%%%%%%%%

The shear catalog is composed of 34.5 million sources, and is nearly identical to that used in \citet{sjs2009}: see that work for further details of the catalog.
The source redshift distribution is obtained by stacking the posterior
probability distribution of the photometric
redshift for each source, $P(\zs)$. Its peak is at $z \sim 0.35$, and
it has a substantial tail extending out to higher redshifts. The full distribution is shown in \citet{cjt2014}, which uses precisely the same source catalog.

In what follows, we describe our lensing
measurement procedure. 
Following the method in \citet{msb2013},
we use, as the lensing observable,
the stacked surface mass
density field at the radial distance $R$ in 
the region around each void,
estimated from the measured shapes of 
background
galaxies as
\begin{equation}
 \Delta\Sigma_{k}(R;\zl)=\frac{\sum_j 
\left[
w_j \left(\ave{\Sigma_{\rm
  crit}^{-1}}_j(\zl)\right)^{-1} \gamma_{k}(R)
\right]
}{\sum_j w_j}
\label{eq:dSigma}
\end{equation}
where 
the summation $\sum_j$ runs over all the background galaxies
in the radial bin $R$, around all the void centers,
the $k$ indices denote the two components of shear (tangential or cross), 
and the weight for the $j$-th galaxy is given by
\be \label{eq:weight}
w_j = \frac{\left[
\ave{\Sigma_{{\rm crit}}^{-1}}_j(\zl)\right]^{2}}
{\sigma_{\rm shape}^2 + \sigma_{{\rm meas},j}^2}.
\ee
We use $\sigma_{\rm shape} = 0.32$ for
the typical intrinsic ellipticities  and 
$\sigma_{{\rm meas},j}$ denotes
measurement noise on each background
galaxy ellipticity. 
$\ave{\Sigma_{\rm crit}^{-1}}_j$ is the lensing critical density for the
$j$-th source galaxy, computed by taking 
into account the photometric redshift uncertainty:
\be
\ave{\Sigma_{{\rm crit}}^{-1}}_j(\zl)
= \int_0^\infty\! {\rm d}\zs \Sigma_{\rm crit}^{-1} (\zl, \zs) P_j(\zs),
\ee
where $\zl$ is the redshift of the void and $P_j(\zs)$
is the probability distribution of photometric redshift
 for the $j$-th galaxy.
Note that $\Sigma_{\rm crit}^{-1}(\zl,\zs)$ is computed as a function of lens
and source redshifts for the assumed cosmology as
\be
\Sigma_{\rm crit} (\zl, \zs) = \frac{c^2}{4\pi G} \frac{D_A(\zs) (1+\zl)^{-2}}{D_A(\zl) D_A(\zl,\zs)} \, ,
\ee
where the $(1+\zl)^{-2}$ factor is due to our use of comoving coordinates, and we set $\Sigma_{\rm
crit}^{-1}(\zl,\zs)=0$ for $\zs<\zl$ in the computation.

%%%%%%%%%%%%%%%%%%%%
\subsection{Jackknife Realizations}
%%%%%%%%%%%%%%%%%%%%

We divide the voids into 128 spatial patches, and perform the measurement multiple times with each region omitted in turn to make $N=128$ jackknife realizations. Note that we exclude the three low-DEC SDSS stripes from our analysis: they are sub-optimal for void finding due to a high ratio of perimeter to area. The remaining area is approximately 7,000 square degrees. The covariance of the measurement \citep{nbg2009} is given by
\begin{flalign} \label{eq:cov}
C [\Delta \Sigma_i, & \Delta\Sigma_j] = \frac{(N-1)}{N}
& 
\nonumber \\
& 
\times \sum\limits_{k=1}^N \left[(\Delta \Sigma_i)^{k} - \overline{\Delta\Sigma_i}\right]
\left[(\Delta \Sigma_j)^{k} - \overline{\Delta\Sigma_j}\right]
\end{flalign}
where the mean value is
\be \label{eq:avg}
\overline{\Delta\Sigma_i} = 
\frac{1}{N}
\sum\limits_{k=1}^N (\Delta\Sigma_i)^k\, ,
\ee
and 
$(\Delta\Sigma_i)^k$
denotes the measurement from the $k$-th realization and the $i$-th
spatial bin. The covariance is measured for both components of shear;
for clarity we do not denote the separate components in Eqs.~\ref{eq:cov} and \ref{eq:avg}.

%%%%%%%%%%%%%%%%%%%
\begin{figure*}
\centering
\resizebox{58mm}{!}{\includegraphics{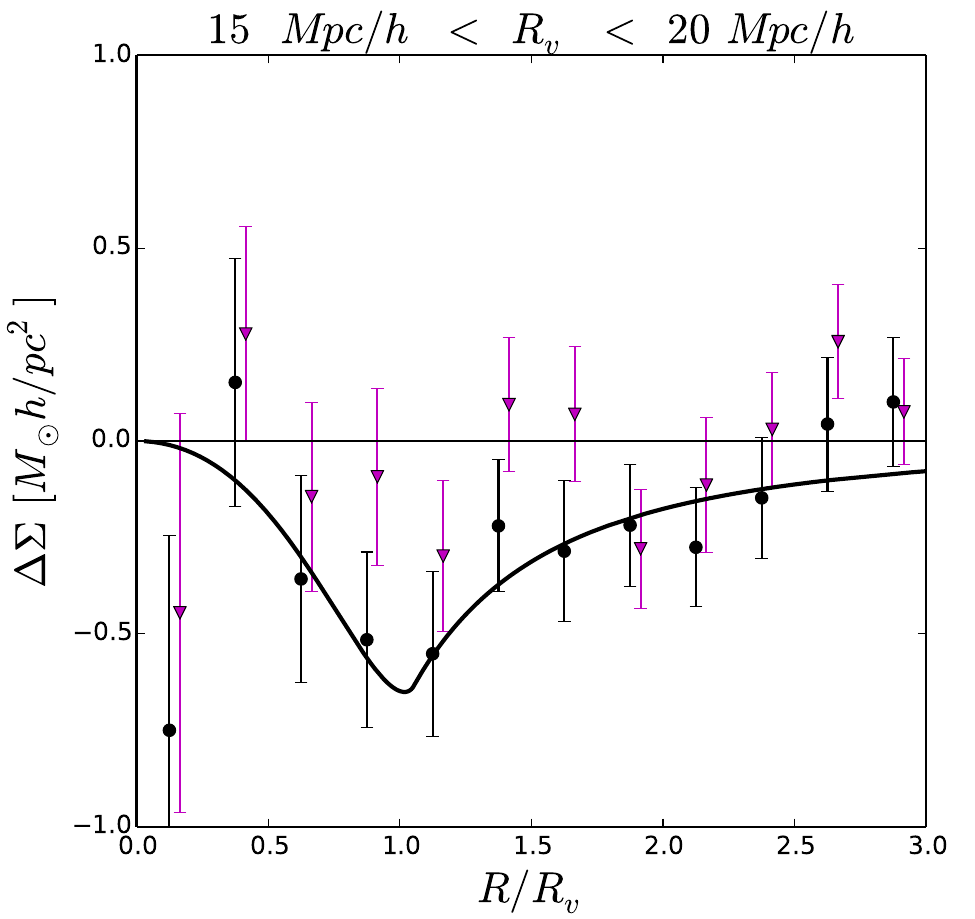}}
\resizebox{58mm}{!}{\includegraphics{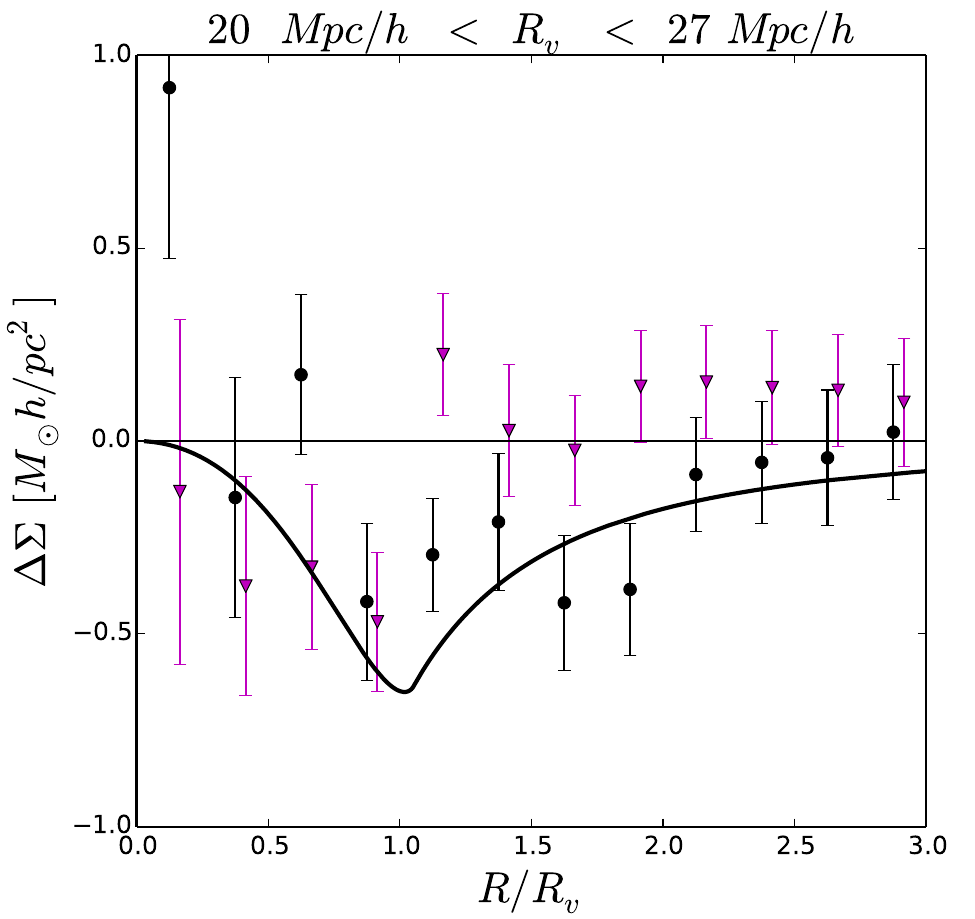}}
\resizebox{58mm}{!}{\includegraphics{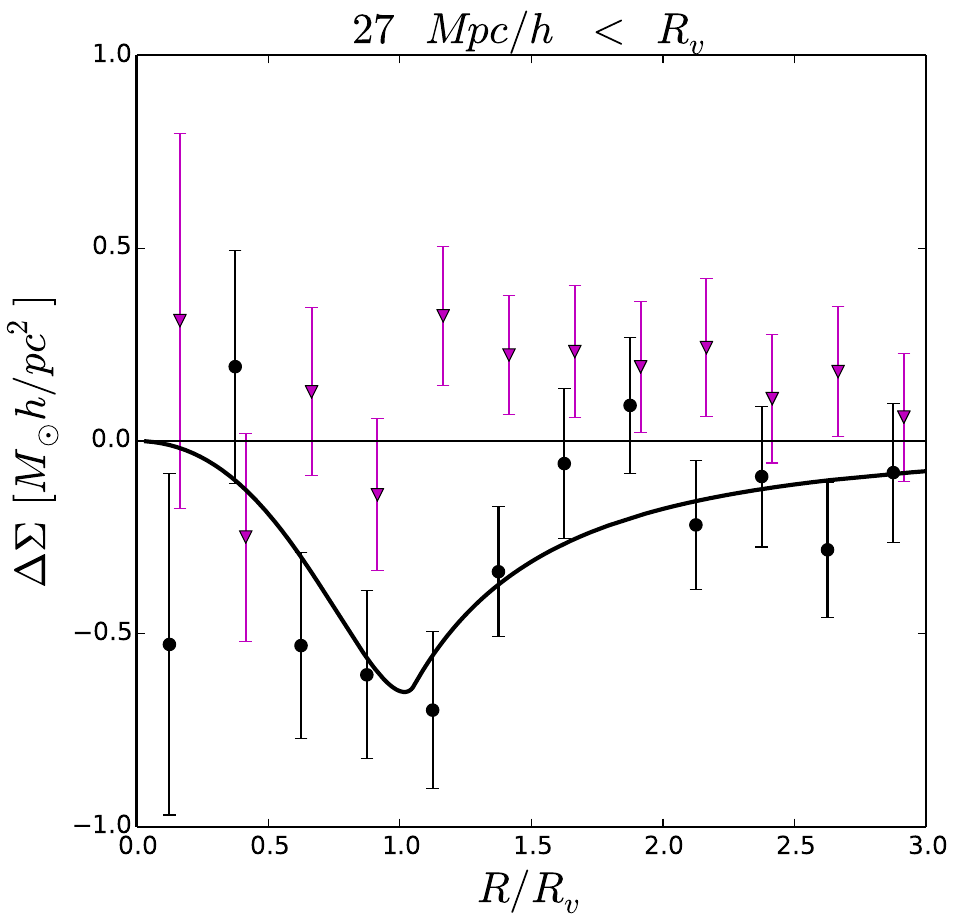}}
\caption{We divide our fiducial measurement (Fig.~\ref{fig:gammat}) into several bins, in order to study the dependence of lensing signal on void size $\rv$. The signal is clearly consistent over a wide range of void sizes, which is a useful test. It has no significant trend with $\rv$. This may be due to the small number of voids above $\sim 30 \mpch$, as well as their large covariance between bins.}
\label{fig:gammat-size}
\end{figure*}
%%%%%%%%%%%%%%%%%%%

%%%%%%%%%%%%%%%%%%%
\subsection{Null tests}
%%%%%%%%%%%%%%%%%%%

We measure the cross-component around void centers and tangential shear around random points, both of which should be consistent with the null hypothesis. With $N$ bins and no model parameters ($n=0$), the null has expected chi-square
\be
\langle \chi^2 \rangle = N - n \pm \sqrt{2N + 2n} \, ,
\ee
so that the expected reduced chi-square is then
\be
\langle \chi^2_{\rm red} \rangle = \frac{\langle \chi^2 \rangle}{N} \sim 1 \, .
\ee
The cross-component is shown in Fig.~\ref{fig:gammat} (pink triangles), and with $\chi_{\rm red}^2 = 22.9/21$ it is easily consistent with the null hypothesis.

We perform the random points test by giving each void with radius $\rv$ and redshift $z$ a random location in the survey area, avoiding masked regions in the same way as the LRG catalog. Often tests involving random points use many more random points than lens galaxies, but since void lenses are so large and many source galaxies fall in each radial bin, we only use as many random points as we have void positions. The result for the tangential shear around random points is $\chi_{\rm red}^2 = 23.7/21$, consistent with the null hypothesis.

%%%%%%%%%%%%%%%%%%%
\subsection{Tangential shear profile}
%%%%%%%%%%%%%%%%%%%

We show the stacked lensing profile of the voids in the left panel of Fig.~\ref{fig:gammat}. The most significant and largest amplitude $\Delta\Sigma$ values of $\sim -0.7 M_\odot h / {\rm pc}^2$ occur at the void radius $\rv$. The signal remains significant out to $\sim 2-3 \rv$. The covariance, shown in Fig.~\ref{fig:cov-gammat}, is used to calculate the signal-to-noise as
\be \label{eq:s2n}
(S/N)^2 = \chi^2 - N_{\rm bin} = \sum_{i,j} \Delta\Sigma_i C^{-1}_{ij} \Delta\Sigma_j - N_{\rm bin} \, ,
\ee
which is the formula for the expectation value of a noncentral chi-squared distribution.
The result is $S/N = 7$ for the measurement in Fig.~\ref{fig:gammat}.
This high significance detection is further supported by the null tests described above.
Note that the number of bins $N_{\rm bin}$ must be subtracted from the $\chi^2$ as in Eq.~(\ref{eq:s2n}) in order to obtain an unbiased measurement of the signal-to-noise.

This result for the signal-to-noise is not very sensitive to the precise values of the cuts described in Sec.~\ref{sec:volume}.
For example, changing the axis ratio cut to 2 or 4 (compared with our fiducial choice of 3) results in a 5\% or smaller change in S/N.
Allowing voids as small as $10\rv$ into the sample increases the signal-to-noise to $\sim 8$, while removing the few hundred voids above $40\mpch$ (see Fig.~\ref{fig:analyze}) causes a negligible change to the results.
As a further check, we compare our measured statistical significance with a rough analytical estimate of the signal-to-noise below.
We then discuss the implications for void density profiles. 

The covariance matrix is largely diagonal up to 1.5 $\rv$. At large $R$ the off diagonal elements are mostly positive, presumably since multiple projections of source galaxies provide less independent information about the voids. In Fig.~\ref{fig:gammat-size} we show three size bins. No systematic trend in magnitude or shape of the signal is visible from these plots. The consistency of the signal across size bins that span nearly a factor of four in void radius validates the lensing interpretation.

%%%%%%%%%%%%%%%%%%%
\subsection{Analytical signal-to-noise estimate}
\label{sec:s2n}
%%%%%%%%%%%%%%%%%%%

We present two checks of our measurement: an analytical estimate of signal-to-noise for void lensing as well as a comparison to the signal-to-noise in SDSS galaxy-galaxy lensing. The tangential shear around a void is given by
\begin{equation}
\gamma_t = \frac{\Delta \Sigma}{\Sigma_{\rm crit}} = \frac{\Sigma(<R) - \Sigma(R)}{\Sigma_{\rm crit}}
\end{equation}
where $\Sigma_{\rm crit}$ is defined above and is  $\Sigma_{\rm crit}
\approx 6000 M_\odot/pc^2$ for our typical lens and source
redshifts. Inside the void radius the signal can be anticipated using
the results of Krause et al (2013):   $\Delta \Sigma \approx -0.6
M_\odot/pc^2$ (adjusted for the fact that our mean void radius is
larger than the range considered in Krause et al). Hence the typical tangential shear
is $\gamma_t \approx 10^{-4}$.

Since our voids and therefore source galaxies are at high redshift, shape and measurement noise both contribute to the errors. We take the noise on the {\it shear} of a given background source to be $\sigma = \sqrt{\sigma_{\rm shape}^2 + \sigma_{\rm m}^2} \sim 0.3$. With a source number density $n
\approx 0.5/{\rm arcmin}^2$, we can then estimate the noise contribution on
a stacked void lensing measurement. For $N_v$ voids of radius
$\theta_v$, we get a sky coverage that exceeds $N_v \pi (2\theta_v)^2$ since the
signal is measured out to at least twice the void radius. This gives a total
effective number of sources $N_{\rm source} = n  N_v \pi (2
\theta_v)^2 \approx 10^9$.
This is thirty times larger than the actual number of source galaxies since each galaxy shape is used
multiple times: it is projected along different directions for different void centers.
Note that this is analogous to galaxy-galaxy lensing, where a given source galaxy contributes to the
density profile of multiple lens galaxies since its shape is projected along different directions for different lens centers.
The volume overlap for galaxy-galaxy lensing exceeds ours for the scales of interest.
The estimated shape noise is then $\sigma_{\rm shape}/\sqrt{N_{\rm source}} \approx  1 \times 10^{-5}$.
The estimated signal to noise is:
\begin{equation}
{\rm S/N} \approx 10.
\end{equation}
Note that in this estimate we have used typical numbers for the void radius $\theta_v = 100$ arcminutes and lensing signal $\gamma_t \approx 10^{-4}$.
Both actually cover a range of values, so there is some uncertainty in our S/N estimate above.
We have also neglected the covariance due to multiple projections of each source galaxy, which could lower the signal-to-noise further.
Given these uncertainties, we are satisfied with the consistency between the analytical estimate above and our  
our measured S/N of 7.

While the estimate above involves several approximations, it gives us a reality check on our measurement. One might still worry that shears at the $10^{-4}$ level are dominated by systematic errors. Indeed for shear-shear correlations from SDSS, that appears to be the case due to additive systematics that are spatially correlated. Such terms however cancel out of cross-correlations. Published measurements of galaxy-galaxy lensing demonstrate this: at distances greater than 10 Mpc the signal falls below $10^{-4}$, see e.g. Figure 6 in Mandelbaum et al (2013). We have checked that the signal-to-noise of that measurement is consistent with ours, adjusting for the smaller number of source galaxies in their  angular bin. Of course closer to the center the galaxy halo overdensity far exceeds the amplitude of the void underdensity, so integrated over all scales the significance of the galaxy-galaxy lensing measurement is higher.

%%%%%%%%%%%%%%%%%%%
\subsection{Comparison with other work}
%%%%%%%%%%%%%%%%%%%

The strength of our detection may be surprising given other work on void lensing. In particular, \citet{mss2013} used a conservative sample of a relatively conservative void finder \citep{slw2012} which was not optimized for lensing purposes. All these factors make a difference in the potential S/N:

\begin{itemize}
\item \citet{mss2013} used the ``central'' sample of \citet{slw2012}; for {\it lrgdim} (the sample most comparable to ours) the usable volume is only 75\% of the total. Furthermore, the void volume fraction is less in the central sample than in the total, where both volume fractions are calculated with respect to their own usable volumes. We make a related quality cut, but which only removes $\sim 1$\% of our sample (Fig.~\ref{fig:analyze}).
\item Over most of the volume where \citet{slw2012} can compare with \citet{pvh2012}, the former finds only half as many voids. This is for the main SDSS galaxy sample, but it is indicative of a difference in void finder aggressiveness between the two methods.
\item Another point worth noting is that our assignment of void radii on the sky is optimized for lensing by setting $\rv$ to the distance to the LRG ridge in the plane of the sky. \citet{slw2012} starts with the void volume and then assign the void radius as $R_{\rm eff} = (3V/4\pi)^{1/3}$, which is used by \citet{mss2013} to bin the background shears. Converting in this way from volume to an effective void radius assumes all three dimensions are the same, but for lensing purposes the line-of-sight size of the void is much less important than its size on the sky. We have tested the effect of assigning an $R_{\rm eff}$ as described above to each of the voids in our fiducial sample and then remeasuring $\Delta \Sigma$ binned in $R/R_{\rm eff}$. The result is an increase in our errors such that the signal-to-noise drops from 7 to 5.
\end{itemize}

%%%%%%%%%%%%%%%%%%%%%%%%%%%%
\section{Void density profile}
%%%%%%%%%%%%%%%%%%%%%%%%%%%%

%%%%%%%%%%%%%%%%
\subsection{Model constraints}
\label{sec:model}
%%%%%%%%%%%%%%%%

The 3-dimensional density profiles of voids have been studied using simulations and other theoretical approaches. One of the subtle issues is how to transition from the underdensity of the void to the cosmic mean density $\bar\rho$ at a sufficiently large distance from the void center. Typically a small transition zone outside the void radius allows for some degree of compensation of the profile, i.e., a region of density higher than $\bar\rho$. In perfectly compensated voids models, the enclosed mass at about two times the void radius is exactly the same as the mass enclosed in a region of the same size with constant density $\bar\rho$.
 
\begin{figure}
\centering
\resizebox{85mm}{!}{\includegraphics{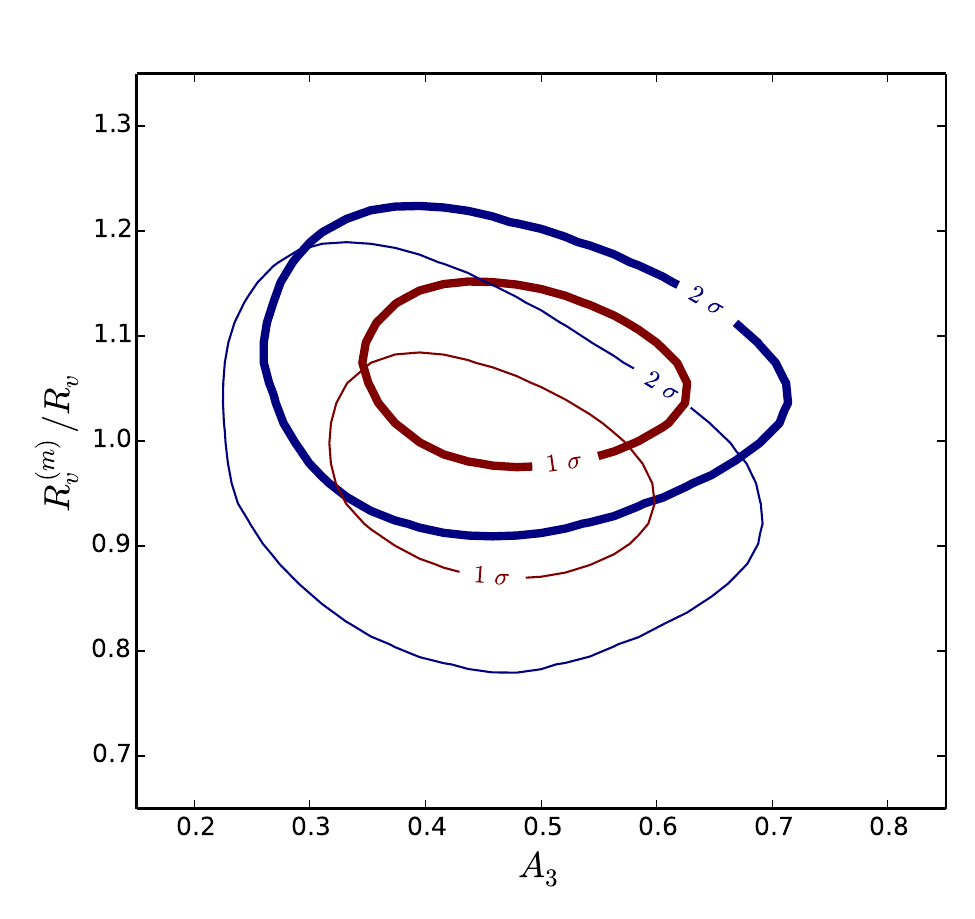}}
\caption{ Thick contours show the 1- and 2-$\sigma$ constraints on our model parameters.
$\rvm$, the radius at which the density profile transitions from cubic to constant, is constrained to be within 10\% of the LRG ridge determined from the data. The constraints for the case requiring volume overlaps of less than 50\% (thin contours) are consistent with our fiducial model (thick contours).}
\label{fig:contour}
\end{figure}

\citet{lw2012} fit a cubic profile inside the void radius using simulations, and \citet{kcd2013} gives the lensing prediction for this model. We use the cubic profile up to the void radius, but outside the void we use a constant density profile. Thus we require continuity at the void radius but not exact compensation. The resulting profile is given by
\be
\rho(r, \rv) = \left\{
\begin{matrix}
\bar{\rho} [A_0 + A_3 (r / \rvm)^3]  \ {\rm for} \ 0 < r < \rvm \cr
\bar{\rho} [A_0 + A_3] \qquad  {\rm for} \ \rvm < r
\end{matrix}
\right. \, ,
\ee
where $A_0, A_3$, and $\rvm / \rv$ are model parameters. However, we are not sensitive to $A_0$, and so have assumed its value is set by requiring that the 3d density returns to the cosmic mean density outside the void, thus $A_0 = 1 - A_3$. Then our fit just involves two parameters, $A_3$ and $\rvm / \rv$, which are constrained as in Fig.~\ref{fig:contour}.
The right panel of Fig.~\ref{fig:gammat} shows the corresponding 3d density profile for our best-fit parameters.

There are a number of differences between our voids and those from which this functional form of the void profile was derived: the simulated voids of \citet{lw2012} are (i) somewhat smaller ($\sim 10 \mpch$ in radius), (ii) traced by dark matter particles instead of halos, and (iii) found using a different void finder.
Despite these differences the fit is acceptable, with a reduced chi-square for our best-fit model of $\chi_{\rm red}^2 = 26.8/19$.
This number is slightly high, but only just outside the $1\sigma$ expectation for a chi-square distribution with 19 degrees of freedom.
This is a reassuring check that the shape of our void profile is at least in qualitative agreement with the previous theoretical work on void profiles by \citet{kcd2013} and \citet{lw2012}.
We leave a more thorough comparison involving simulations that match our void properties carefully for future work.

%%%%%%%%%%%%%%%%%%%
%%%%%%%%%%%%%%%%%%%

If we were to require compensation, as in some models explored by \citet{kcd2013}, we would put some constraints on $A_0$. Note also that \citet{hoh2013} apply the \citet{hv2002} void finder to ray-traced simulations, and their tangential shear profiles look compensated just beyond the void radius.
Assuming that $\Sigma = \bar{\Sigma}$ by $2\rv$ (see below), the data prefers an uncompensated void inside $2\rv$. This is shown by the negative values of the measured $\Delta\Sigma$ up to and beyond $2\rv$ (it should be zero for a compensated void if $\Sigma(2\rv) \rightarrow \bar{\Sigma}$).
However, some caution is required in interpreting the best-fit model pictured in the right panel of Fig.~\ref{fig:gammat}.
First, note that $\Delta\Sigma$ is a nonlocal quantity, such that data points even at $\sim 2-3 \rv$ can be dominated by the density profile well inside $\rv$.
This makes direct comparisons of the left and right panel of Fig.~\ref{fig:gammat} difficult.
Second, while the best-fit model value is underdense at $\rv$, there is a substantial error bar which shows that densities at $\rv$ larger than the cosmic mean are not ruled out.
Furthermore, given the errors on the measurement, this model is not the only one that would provide an adequate fit to the data.
More work is needed to understand the relationship of the LRGs to the mass profile as  we expect that our void finder played some role in the details of the LRG profile. 

While the minimum density at the center of the void is formally not constrained by the data, we find that  the requirement that the density approach the mean at large radii, coupled with measurements between $\rv$ and $2\rv$, leave little freedom. We explored modifications to the density profile beyond $\rv$ and find that $A0$ can be lowered by at most $0.1$. The arrow in Fig.~\ref{fig:gammat} (right panel) pointing to lower central densities indicates this possibility. 

Also shown on Fig.~\ref{fig:contour} is the effect of stricter criteria for void overlap, for the case with overlaps well below 50\%. The constraints are degraded very little due to throwing away a large fraction of overlapping voids, and the shift in the contours is negligible for $A_3$ and less than $1\sigma$ for $\rvm$.

%%%%%%%%%%%%%%%%%%%%
\subsection{Estimated mass deficit inside the voids}
%%%%%%%%%%%%%%%%%%%%

Since the measured $\Delta\Sigma = \Sigma(<R) - \Sigma(R)$, we can estimate $\Sigma(<\rv)$  once we require $\Sigma$ to approach $\bar{\Sigma}$ at some large radius. At radii above  $2\rv$ both the galaxy distribution and the mass in simulations \citep{lw2012} are close to the mean density. These are large scales, typically above $ 40 \mpch$, so it is reasonable to expect that there aren't  departures at more than a few percent level from mean density in the data as well. 
We therefore use our measurements at about $2\rv$ to estimate $\Sigma(<\rv)$ with this assumption. We test it by checking the range $1.5-2.5\rv$, at which our signal to noise is still reasonable. 

The results for the mass deficit and fractional mass deficit are shown in Table 1. Two methods are used: directly from the data as described above and using our best fit for $\rho(r)$.
Each estimate involves some assumptions or caveats which are briefly described in the table. 
The mass deficit 
\begin{equation}
\delta M = \frac{4\pi}{3} \rv^3\left[\rho(<\rv)-\bar{\rho}\right]
\end{equation}
is estimated only for the 3d model fit.

%While we have not attempted to place rigorous bounds on our estimated $\delta M$ values, we can see the trends between data and simulations: the two methods of estimation from the data are in reasonable agreement, and involve more mass inside voids than in simulations (the deficit is about 40\% higher in the simulation fits).
%\red{These simulated voids are somewhat smaller (at $\sim 10 \mpch$ in radius) than ours and were found using a different void finder.

Our measurements indicate significant levels of underdensity inside the void radius: the inferred 3d fractional under density is $\approx -0.3$ to $-0.4$ inside $\rv$. This corresponds to mass deficits comparable to the masses of the most massive clusters in the universe. The bigger voids in our sample will have up to ten times  the mass deficit. Given that our LRG sample has a bias factor of about 2, we expect that voids using a less biased tracer would have lower central densities. Simulations with mock catalogs also support this trend \citep{slw2014}.
We also extended the profile from $\rv$ to $2\rv$ using different models, including a possible ridge of density above the mean, but find that the measurements leave little wiggle room.
Note however that projection effects and flaws in the void finder could lead us to overestimate the mass enclosed.
We leave for future work a detailed comparison with simulations to estimate the net dilution of signal due to such spurious voids.

\begin{table*}
\centerline{
\small
\begin{tabular}{| l | c | c | c | c |}
	\hline
{\bf Method} & $\delta M(<\rv)$ & $\rho(<\rv)/\bar{\rho} -1 $ & $\Sigma(<\rv)/\bar{\Sigma}-1$ & {\bf Assumptions} \\ \hline
Measured $\Delta\Sigma(R_v - 2 R_v)$ & --  & --  & -0.2 (-0.3) & assume $\Sigma(R) \rightarrow \bar{\Sigma}$ at $R\approx \rv (2\rv)$ \\ \hline 
Best fit model & 
$\approx -1 \times 10^{15} M_\odot$ &  -0.3 (-0.4) & -0.22 (-0.32) & Fix $A_0$ to recover mean density at $R\approx \rv (2\rv)$ \\ \hline 
%Fit from simulations & 
%$\approx -1.4\times 10^{15} M_\odot$ &  -0.5 & -0.44 & Different void finder. No projection effects. \\ \hline 
\end{tabular}
}
\caption{\small Estimated mass deficit $\delta M$ and the fractional deficit in the 3d density $\rho$ and projected density $\Sigma$ at the void radius $\rv$.  The measurements, interpreted without a model in the first row, give us only projected quantities. For the model fits we give both 2d and 3d versions of the fractional density contrast. We set $\rv = 20\mpch$ to estimate $\delta M$; for voids with other values of $R_v$, $\delta M$ scales approximately as $R_v^3$.
  }
\end{table*}

%%%%%%%%%%%%%%%%%%%%%%%%%%%%
\section{Discussion}
%%%%%%%%%%%%%%%%%%%%%%%%%%%%

{\it Void Lensing Detection}.
We have made a high signal-to-noise measurement of gravitational lensing by large voids (Fig.~\ref{fig:gammat}), obtaining the first constraints on the dark matter density profile of voids. This measurement may be surprising given that theoretical work \citep{kcd2013} predicted that ambitious future surveys (in particular, Euclid) would be needed for measurements with comparable signal-to-noise. We differ from previous work in that our void finder and void characterization  is optimized for lensing. We work with projected 2d slices    and have a flexible criterion that allows for some overlap between voids. Our stacked shear measurement is analogous to galaxy-galaxy lensing in that it projects a source galaxy shape along multiple void centers. This greatly increases the total number of lens-source pairs and reduces shape noise by a factor of several. Other improvements described in Section 2 contribute to the size and quality of our void sample. 

We validate our detection of void lensing in several ways, using both the LRG positions around voids and standard galaxy-shear tests. Figures \ref{fig:lrg-dens} and \ref{fig:analyze} show the validation and improvements based on the LRG distribution. We verify that the tangential shear around random points and the lensing cross-component around void centers are consistent with the null hypothesis.  The error analysis is analogous to that for our measurement of filament lensing with the same dataset presented in \citet{cjt2014}.

{\it Void density profiles}.
We measure the stacked density profile of voids with radii $R_v = 15-55 \mpch$ over the redshift range $0.16 < z < 0.37$. We can make some model-independent statements about void properties (see Table 1). By requiring the projected density to approach  the mean density at radii of $2\rv$ or larger, we can convert our measured $\Delta\Sigma$ to estimates of $\Sigma(<\rv)$ and therefore to the fractional density contrast at $\rv$. We further estimate the mass deficit $\delta M$. In addition, we find that our voids are uncompensated within twice the void radius. By $3\rv$ however, the measurements are consistent with fully compensated voids, but we see no evidence for overcompensated voids of the kind seen in simulations (at the lower end of our $\rv$ range). 

By fitting our measurements with a model motivated by simulations, we can draw conclusions about the voids' 3d density profile and mass deficit $\delta M$ as summarized in Table 1. Our data is consistent with a central density of $\approx 0.5 \bar{\rho}$. At the edge of the void, it is also consistent with a density below the mean density at the LRG ridge, though the corresponding 2d density of LRGs is above the mean (Fig.~\ref{fig:lrg-dens}).

{\it Caveats}.
The standard disclaimer with void-related work is that the results can be quite sensitive to the specific void-finder used. As highlighted above, this holds true also for our work which is designed to find voids for gravitational lensing. Our use of multiple potential void centers is helpful for lensing S/N reasons, but also makes interpretation of the resulting density profile less straightforward. We expect some miscentering between the lowest dark matter density and the emptiest places in the sparse galaxy density, and our multiple centers may also add to this miscentering in some instances.
However, we have checked the results with much stricter criteria for the number of multiple centers, and find only small shifts in the parameter contours (Fig.~\ref{fig:contour}).
Furthermore, since the density profiles are very flat between the center and half the void radius, these effects are far less problematic than for galaxy or cluster lensing.

We expect our error bars accurately account for shape noise and sample variance. However, we have not accounted for possible shear calibration errors, which could bias the signal by up to $ 5\%$. In addition, two effects could result in a dilution of the signal and thus underestimation of $A_3$: inaccurate source redshifts or fake voids from chance LRG projections. We have not estimated the contribution of these effects. 

{\it Future Work}.
We can attempt a void lensing measurement with several different variants of the void sample. Going beyond our sparse sample of LRGs, we can apply this void finder to the SDSS Main sample. Although the volume probed will be significantly smaller, this disadvantage is offset in part by the larger number of background sources available behind lower redshift voids. Furthermore, \citet{slw2014} find that the voids identified using a lower galaxy luminosity threshold have a lower central dark matter density (as expected based on their lower galaxy bias as well), which should increase the lensing effect. 

Nearly all detailed applications will require a careful study of our void selection via mock catalogs that create galaxies from HOD prescriptions or dark matter halos. Our measurements are now confined to $\rv > 15 \mpch$, in part because the contamination from fake voids due to projection effects gets worse as the void size gets smaller than the 2d tracer density. Mock catalogs will allow us to go down to smaller radii and estimate the number of fake and real small voids. With those numbers we can take into account the expected dilution of the signal. 

The comparison of the galaxy distribution with the mass distribution is of great interest. The question of galaxy biasing can be understood better by having measurements in under dense regions to complement  those in over dense regions. Many other questions can be posed by stacking voids in different ways: along the major axis of the galaxy distribution, varying the environment and the properties of the galaxy population, and so on. The measurement of a magnification signal behind voids would be of interest, in particular to provide a direct measurement of $\Sigma(R)$.

Void mass functions, mass  profiles, and the cross-correlation with  galaxy profiles are the key ingredients in cosmological applications of voids. The velocity profiles measured in SDSS have an anisotropy and relationship to the mass profile that carry cosmological information \citep{lw2012}.  Modified gravity theories in particular predict differences in these observables. In many respects modeling voids is less problematic than massive nonlinear objects like galaxy clusters, and the measurements are not affected by  foreground galaxies, but the use of mock catalogs to understand the selection effects in the data is likely to be essential to interpreting survey measurements. 

%%%%%%%%%%%%%%%%%%%
\section*{Acknowledgments}
We would like to thank Gary Bernstein, Sarah Bridle,  Doug Clowe, Daniel Gruen, Nico Hamaus, Adam Lidz, Hironao Miyatake, Ravi Sheth, Paul Sutter, and Vinu Vikram for helpful discussions. Mike Jarvis and Elisabeth Krause gave us many insightful suggestions as well as valuable feedback on the paper. Robert Lupton's comments on the void finder led to a much more concise and intuitive description of the algorithm. We are very grateful to Erin Sheldon for the use of his SDSS shear catalogs.
Some of the results in this paper have been derived using the HEALPix \citep{ghb05} package.
BJ and JC are partially supported by Department of 
Energy grant DE-SC0007901.

%%%%%%%%%%%%%%%%%%%
% Bibliography
%%%%%%%%%%%%%%%%%%%

%%%%%%%%%%%%%%%%%%%

\end{document}